\documentclass[twocolumn,preprintnumbers,amsmath,amssymb,showpacs,prl]{revtex4}

\usepackage{graphicx}
\begin{document}


\title{Non-Photonic Electron $p_{T}$ Distributions and Correlations of Electrons from $B$ and $D$ Meson Decays with Charged Hadrons}


\author{Xiaoyan Lin\\Institute of Particle Physics, Central China Normal University, Wuhan 430079, China\\
Department of Physics and Astronomy, University of California, Los Angeles, CA 90095, USA}

\date{\today}

\begin{abstract}
We compare the non-photonic electron and $D^{0}$ meson $p_{T}$
distributions measured at RHIC with the PYTHIA Monte Carlo event
generator in p+p collisions at $\sqrt{s_{NN}}=200$ GeV. A delta
fragmentation function much harder than the Peterson function,
consistent with recombination scheme for charm meson formation, is
needed to match the experimental data. Attempts to fit the
non-photonic electrons at high $p_{T}$ show large uncertainties and
the $B$ meson semi-leptonic decays may not be dominant for electron
$p_{T}$ up to 8 GeV/c. Correlations of non-photonic electrons with
charged hadrons are studied. We propose an experimental method to
quantitatively determine the relative contributions of $D$ and $B$
meson semi-leptonic decays to the non-photonic electrons.

\end{abstract}


\maketitle

Heavy quarks are believed to be produced through initial
parton-parton, mostly gluon-gluon, scatterings in nuclear collisions
at the Relativistic Heavy Ion Collider (RHIC) energies. Theoretical
calculations of heavy quark production within the perturbative
Quantum ChromoDynamics (pQCD) framework are considered more reliable
because the heavy quark mass sets a natural scale for the pQCD.
Charm quark production is sensitive to the incoming parton flux for
the initial conditions of nuclear
collisions~\cite{appel}\cite{mull}. The transport dynamics of the
heavy quarks in nuclear medium such as flow~\cite{rapp} and energy
loss~\cite{doks}\cite{bw-zhang} can probe QCD properties of the
dense matter created in nucleus-nucleus collisions. Therefore, heavy
quark measurements provide unique insights into QCD properties of
the new state of matter produced in nucleus-nucleus collisions.

The heavy quark production in p+p or p+A collisions provides a
reference for heavy meson formation and for nuclear modification
factors of heavy quarks in the nuclear medium. The Peterson
function~\cite{peterson}
\begin{eqnarray}
\label{eq1} D(z {\equiv} p_{D}/p_{c}) {\propto} \frac
{1}{z[1-1/z-{\varepsilon}/(1-z)]^2}
\end{eqnarray}
has often been used to describe the charm fragmentation function,
where the parameter $\epsilon \approx 0.05$ (default in
PYTHIA~\cite{pythia}) is in reasonable agreement with the results
from fits to charm production in $e^+e^-$ and $\gamma p$
collisions~\cite{cfrag}. However, in charm hadroproduction, it was
observed that the $c$-quark $p_{T}$ distributions of
next-to-leading-order (NLO) pQCD calculations agree well with the
measured open charm $p_{T}$ spectrum~\cite{barequark}, indicating
that a much harder fragmentation function peaked at $z \approx 1$ in
Eq.~\ref{eq1} is needed in charm hadroproduction. A more detailed
discussion of this observation can be found
in~\cite{vogt1992}\cite{cbigpaper}.
Coalesence~\cite{ZW-lin}\cite{lin} or
recombination~\cite{duke}\cite{hwa}\cite{greco} models have also
been proposed for charmed meson formation by combining a charm quark
with a light up or down quark, presumably of soft $p_T$~\cite{coal}.
Thus the charmed hadron $p_T$ would coincide with the bare charm
quark $p_T$ distribution in this hadronization scheme. These various
hadron formation schemes can lead to significantly different charmed
meson $p_T$ distribution when interpreting non-photonic electron
$p_T$ spectra from experimental measurements.

Recently STAR and PHENIX collaborations have observed a suppression
of non-photonic electrons much larger than predicted in the $p_{T}
\sim 4-8$ GeV/c region~\cite{star1}\cite{phenix1}. This observation
challenges theoretical predictions based on energy loss via induced
gluon radiation for heavy quarks~\cite{armesto}\cite{djordjevic}.
The fact that the relative contributions of $D$ and $B$ meson
semi-leptonic decays to the experimental non-photonic electron
spectrum is currently not measured at RHIC severely restricts the
understanding of the heavy quark energy loss. A nonzero non-photonic
electron $v_{2}$ has been measured in the $p_{T} < 2.0$ GeV/c
region, while at higher $p_{T}$ the $v_{2}$ is observed to decrease
with $p_T$~\cite{akiba}. Quantitative understanding of this feature
in heavy quark measurements requires relative charm and bottom
contributions to non-photonic electrons.

In this paper we evaluate the $p_{T}$ distribution of $D$ mesons
from PYTHIA v6.22~\cite{pythia} and compare the PYTHIA results with
the STAR measurements~\cite{adams}\cite{an}. The charm quark
fragmentation function will be modified from the default Peterson
function and the other PYTHIA parameters are tuned in order to
describe the experimental $D$ meson data as well as the STAR
non-photonic electron spectra~\cite{star1}. We study the
correlations between non-photonic electrons and charged hadrons in
order to guide the experimental investigations of the relative
contributions to the non-photonic electron spectrum at high $p_{T}$
from $B$ and $D$ meson decays.

\begin{figure}
\includegraphics[width=3.5in]{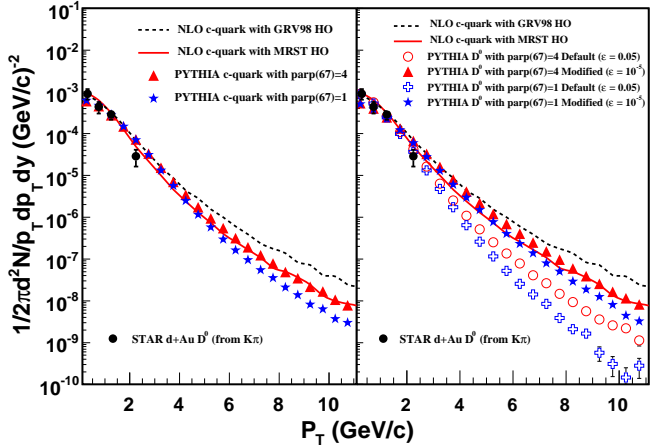}
\caption{\label{fig:D0}(color online) Charm quark (left) and $D^{0}$
(right) spectra from PYTHIA calculations compared with
next-to-leading-order pQCD predictions for charm quark spectra and
the STAR measured $D^{0}$ spectrum from d+Au collisions scaled by
$N_{bin}=7.5$.}
\end{figure}

Fig.~\ref{fig:D0} shows charm quark and $D^{0}$ meson spectra from
PYTHIA calculations and NLO pQCD predictions for charm quark
spectra~\cite{vogt2003} together with the STAR $D^{0}$
spectrum~\cite{adams}, where the measured $D^{0}$ data points from
d+Au collisions at $\sqrt{s_{NN}}=200$ GeV have been scaled by
$N_{bin}=7.5$ -- the number of binary collisions. The PYTHIA spectra
have been scaled to the measured $dN/dy$ of $0.028\pm 0.004$({\it
stat.})$\pm 0.008$({\it syst.})~\cite{adams} for $D^0$ at
mid-rapidity, where the scaling factors are 1.75 for $D^{0}$ meson
spectra and 1.25 for charm quark spectra, respectively. The
theoretical curve with the MRST HO PDF has been normalized to the
measured total $c\overline{c}$ cross section ($1.3\pm 0.2\pm 0.4
mb$)~\cite{adams} by a factor of 3.4.

The stars in the left panel of Fig.~\ref{fig:D0} depict the charm
quark $p_{T}$ distribution from a PYTHIA calculation with the
following parameters (set I) from reference~\cite{adcox}: PARP(67)
$=$ 1 (factor multiplied to $Q^2$) , $<k_t> = 1.5 GeV/c$, $m_c =
1.25 GeV/c^2$, $K_{factor}$ = 3.5, MSTP(33) $=$ 1 (inclusion of $K$
factors), MSTP(32) $=$ 4 ($Q^2$ scale) and CTEQ5L PDF. We further
tuned the value of PARP(67) to 4, which enhances $c$-quark
production probability through gluon splitting and is introduced to
take into account higher order effects in the pQCD
calculation~\cite{norrbin}. The results are shown as triangles in
the left panel of Fig.~\ref{fig:D0}. We found that this change
mainly affects $c$-quark production at high $p_{T}$ and can
effectively reproduce the NLO pQCD calculation. We will refer to
this set of parameters as parameter set II in the rest of the paper.

\begin{figure}
\includegraphics[width=3.5in]{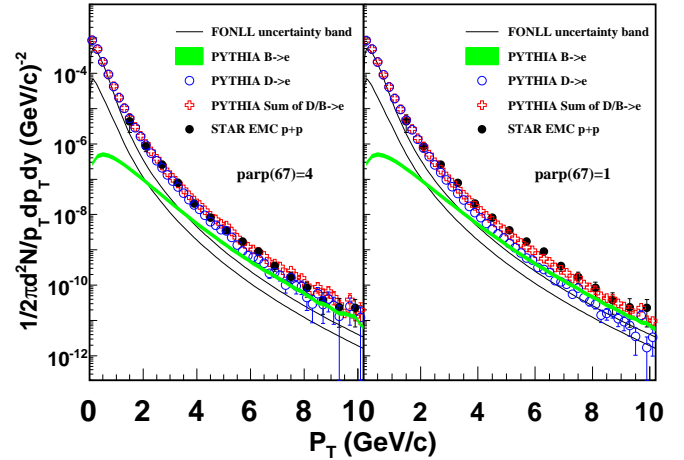}
\caption{\label{fig:elec}(color online) Electron spectra from PYTHIA
calculations with the modified Peterson function for charm quarks
and beauty quarks compared with background subtracted single
electron spectrum measured by STAR from p+p collisions and the
prediction for theoretical uncertainty electron band. Left panel:
The spectra are from parameter set II. Right panel: The spectra are
from parameter set I.}
\end{figure}

We compared the charm quark spectra from PYTHIA calculations to the
STAR published $D^{0}$ spectrum~\cite{adams} together with the STAR
preliminary $D^{*}$ spectrum~\cite{an}. The combination of STAR
measured $D^{*}$ and $D^0$ meson spectra into a single $D$ meson
spectrum is hampered by the large uncertainties in the ratio
$D^{*}/D^0 = 0.4 \pm 0.09 \pm 0.13 $~\cite{an} which is
experimentally not well known at RHIC. The STAR $D$ meson spectrum
covers a limit $p_{T}$ range and is not well constrained at high
$p_{T}$. Without further tuning other PYTHIA parameters, we find
that the generated bare charm quark spectra from PYTHIA calculations
using parameter set I and parameter set II approximately match the
STAR $D$ meson $p_T$ distribution. As demonstrated in the left panel
of Fig.~\ref{fig:D0}, the PYTHIA calculation with parameter set II
also yields a charm quark $p_{T}$ distribution similar to the NLO
pQCD calculations~\cite{vogt2003}\cite{cacc}, which coincides with
the STAR $D$ meson $p_T$ spectrum as well. It is not the purpose of
this paper to show that the PYTHIA calculations with these two
parameter sets and NLO pQCD calculations are equivalent in physics
contents.

In the right panel of Fig.~\ref{fig:D0} the $D^{0}$ $p_{T}$
distribution from PYTHIA calculations using the default Peterson
fragmentation function is shown as open circles for parameter set II
and crosses for parameter set I. The default Peterson function
refers to the value of the parameter ${\varepsilon}$ in Peterson
function being 0.05 for charm quarks and 0.005 for beauty quarks.
The default Peterson fragmentation for charm quarks is too soft to
reproduce the measured $D^{0}$ spectrum~\cite{adams} together with
$D^{*}$ spectrum~\cite{an}. We modified the value of the parameter
${\varepsilon}$ to $10^{-5}$ for both charm and beauty quarks. In
this case the fragmentation function is nearly $\delta(1-z)$. The
results are shown as stars for parameter set I and triangles for
parameter set II in the right panel of Fig.~\ref{fig:D0}. The PYTHIA
calculations using the modified Peterson fragmentation function
(${\varepsilon}=10^{-5}$) with parameter set I and set II can
reasonably reproduce the measured $D$ meson $p_{T}$ distribution.

While the $k_{T}$ broadening can make the charm $p_{T}$ distribution
harder at low beam energies~\cite{cbigpaper}, the intrinsic $k_{T}$
has little effect on the $p_{T}$ distribution at RHIC
energies~\cite{vogt2003}. A harder fragmentation function is needed
for the hadronization of charm quarks if the pQCD calculation is to
describe the measured STAR data.

\begin{figure}
\includegraphics[width=3.5in]{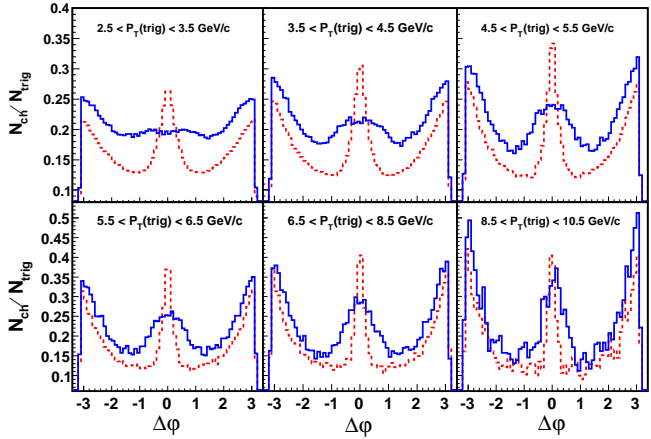} \caption{\label{fig:deltaphi}(color online) $\Delta\varphi$ distributions
between non-photonic electrons and charged hadrons with different
electron trigger $p_{T}$ cuts and associated hadron $p_{T} > 0.1$
GeV/c. Solid lines show electrons from $B$ meson decays. Dashed
lines show electrons from $D$ meson decays.}
\end{figure}

The STAR independent measurements of the reconstructed $D^0$ meson
and single electrons from heavy quark semi-leptonic decays measured
with TOF and TPC are consistent~\cite{adams}. The electron
measurement there only covers up to $p_{T} < 4$ GeV/c and has no
sensitivity to the $B$ meson contribution. We also checked the
consistency between $D$ meson data and non-photonic single electron
data in p+p collisions within our PYTHIA calculations.
Fig.~\ref{fig:elec} shows the electron spectra from PYTHIA
calculations using the modified Peterson fragmentation function for
charm quarks and beauty quarks with parameter set II in the left
panel and parameter set I in the right panel. The parameters for
beauty quarks in set II and set I are all the same as for charm
quarks except $m_b = 4.8 GeV/c^2$. The PYTHIA spectra of electrons
from charm meson decays are scaled by the same factor used to scale
the PYTHIA $D^0$ spectra to the measured $dN/dy(D^0)$ at
mid-rapidity. The electron spectra from beauty meson decays are
normalized by the ratio of $\sigma_{b\bar b}$/$\sigma_{c\bar c}$
based on the NLO pQCD calculation~\cite{vogt2002}. The band
corresponds to the theoretical uncertainty of this ratio (0.45\% -
0.60\%)~\cite{vogt2002}, where the theoretical error may be an
underestimate according to recent calculations~\cite{cacc}. We used
the value at the center of this range to calculate the sum of the
electrons. Fig.~\ref{fig:elec} also shows the comparison to the STAR
measured non-photonic single electron data in p+p
collisions~\cite{star1} and the FONLL calculation for the
theoretical uncertainty band of the electron spectrum from charm and
bottom in p+p collisions~\cite{cacc}. The FONLL prediction of
electron spectrum gives a fair description of the shape of the
measured spectra. But there is a discrepancy in the overall scale
and the FNOLL calculation is significant below the STAR data at high
$p_{T}$. More discussions can be found in~\cite{star1}. The PYTHIA
calculations with parameter set II and modified heavy quark
fragmentation function can simultaneously describe the STAR direct
$D$ meson $p_T$ distribution~\cite{adams}\cite{an} and the STAR EMC
non-photonic electron data~\cite{star1}. This modified scheme
indicates that the contribution of electrons from $B$ meson decays
is not dominant for the measured $p_T$ region up to 8 GeV/c assuming
$\sigma_{b\bar b}$/$\sigma_{c\bar c} \sim 0.45\% - 0.60\%$ based on
NLO pQCD calculation. The measurement of electron $p_T$ distribution
alone has a reduced sensitivity to the $p_T$ distribution of $D$
mesons as shown in reference~\cite{batsouli}. Within the statistical
and systematic errors of the electron data, the PYTHIA calculation
with parameter set I and the modified Peterson function for both $c$
and $b$ quarks can yield an electron $p_T$ distribution similar to
the measurement ~\cite{star1}. In this case, as shown in the bottom
panel of Fig.~\ref{fig:elec}, electrons from $b$ quark decays are
not the dominant source for the $p_T$ region up to 6 GeV/c. It is
critical that the $B$ and $D$ meson decay contributions to
non-photonic electrons to be separated experimentally.

\begin{figure}
\includegraphics[width=3.5in]{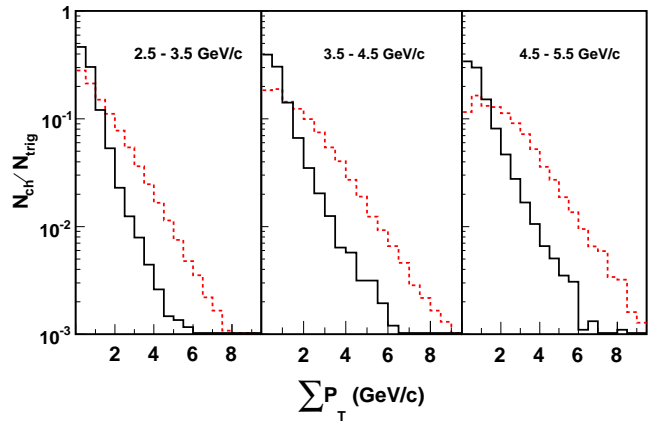} \caption{\label{fig:sumpt}(color online) Summed $p_{\rm T}$
distributions of charged hadrons around triggered non-photonic
electrons with different trigger $p_{T}$ cuts. Solid lines show
electrons from $B$ meson decays. Dashed lines show electrons from
$D$ meson decays.}\end{figure}

We have studied the azimuthal correlations between heavy quark
semi-leptonic decay electrons and inclusive charged hadrons.
Fig.~\ref{fig:deltaphi} shows the $\Delta\varphi$ distributions
between non-photonic electrons and inclusive charged hadrons, with
different electron trigger $p_{T}$ cuts and the associated hadron
$p_{T} > 0.1$ GeV/c. The distributions are scaled by the number of
electron triggers. These plots are from PYTHIA calculation with
parameter set II and the modified heavy quark fragmentation
function. The solid lines in Fig.~\ref{fig:deltaphi} are for
electrons from $B$ meson decays and the dashed lines are for
electrons from $D$ decays. As demonstrated in
Fig.~\ref{fig:deltaphi}, there is a significant difference between
$B$ and $D$ meson decays in the near-side correlations. The width of
near-side peak for electrons from $D$ decays is much narrower than
those for the $B$ decays. The wide width from $B$ meson decays is
due to the larger energy release ($Q$ value) in the $B$ meson
semi-leptonic decays leading to a broad angular correlation between
daughter hadrons and electrons. An electron at high $p_{T}$, if it
is from $B$ meson decays, the $B$ meson does not have to be at high
transverse momentum because the electron can get large momentum from
the $b$-quark mass. In the case of $D$ meson decays, the $D$ meson
needs to have a large momentum in order to boost the daughter
electron to a high $p_{T}$. In experimental effort we cannot
differentiate heavy quark decay hadrons from others, we used
inclusive hadrons in our correlation study. We found the difference
in the near-side correlations between $D$ decays and $B$ decays is
largely due to the decay kinematics, not the production dynamics.
This difference can help us estimate the relative $B$ and $D$
contributions to the yields for non-photonic electrons.

The bottom contribution to the non-photonic electrons can be
determined directly from the electron spectra in PYTHIA. The results
are given in the second column of Table~\ref{table1}. However, this
way to determine the bottom contribution is not experimentally
feasible. In order to experimentally determine the bottom
contribution fraction, we use the $\Delta\varphi$ distributions for
$B$ decays and $D$ decays to fit the $\Delta\varphi$ distribution
for PYTHIA inclusive case (either $B$ decays or $D$ decays), and let
the $B$ contribution fraction as a parameter. The fraction is
determined by the minimum value of $\chi^{2}$. And the fitting error
is determined by one $\sigma$ shift of $\chi^{2}/ndf$ from the
minimum value. The fitting error will reduce by increasing the
statistics. The results are shown in the third column of
Table~\ref{table1}. The results are consistent with those directly
from the electron spectra. This indicates that the method we
proposed is self-consistent. This approach can be used to
experimentally estimate the $B$ and $D$ contributions to the
non-photonic electrons.

\begin{center}
\begin{table}
\caption{\label{table1}Fractions of $B$ meson decay contributions
from electron spectra and $\Delta\varphi$ distribution fitting in
PYTHIA calculations with parameter set II and modified Peterson
fragmentation function for charm and beauty quarks.}
\begin{tabular}{c|c|c}
\hline \hline $p_{T}(trig)$ (GeV/c) &From Spectra ($\%$) &From Fitting ($\%$)\\
\hline 2.5 - 3.5 &$12.64 \pm 0.04$ &$12.64 \pm 1.37$ \\
\hline 3.5 - 4.5 &$21.65 \pm 0.12$ &$21.65 \pm 2.55$ \\
\hline 4.5 - 5.5 &$30.50 \pm 0.26$ &$30.50 \pm 4.76$ \\
\hline 5.5 - 6.5 &$36.66 \pm 0.48$ &$36.66 \pm 7.87$ \\
\hline 6.5 - 8.5 &$42.24 \pm 0.71$ &$42.24 \pm 11.10$ \\
\hline 8.5 - 10.5 &$49.82 \pm 1.72$ &$49.82 \pm 25.11$ \\
\hline
\hline
\end{tabular}
\end{table}
\end{center}
\vskip -1.035cm

We further studied the particle production within a cone around
triggered high $p_{T}$ electrons from heavy quark decays. We focused
on the scalar summed $p_{T}$ distributions of inclusive charged
hadrons in the cone ($p_{T}$ refers to the transverse momentum in
the laboratory frame). Here the cone is defined by $|\eta_{h} -
\eta_{e}| < 0.35$ and $|\varphi_{h} - \varphi_{e}| < 0.35$ ($\eta$
is pseudorapidity and $\varphi$ is azimuthal angle). The summed
$p_{T}$ distributions of inclusive charged hadrons in three
triggered electron $p_{T}$ ranges are shown in Fig.~\ref{fig:sumpt}.
The distributions are scaled to unit. The dashed lines are for $D$
decays and the solid lines are for $B$ decays. These plots are from
PYTHIA calculation with parameter set II and the modified heavy
quark fragmentation function. We also can see that there is a
significant difference between $B$ decays and $D$ decays. The summed
$p_{T}$ distributions for $D$ meson decays are much wider than those
for $B$ meson decays. This difference can also be used to
distinguish $B$ and $D$ decay contributions. We use the summed
$p_{T}$ histograms from $B$ decays and $D$ decays to fit the summed
$p_{T}$ histogram from PYTHIA inclusive case to determine the $B$
decay contribution. The results are shown in Table~\ref{table2}. The
results are consistent with those directly from the electron
spectra. The ratios for the same trigger $p_{T}$ ranges are
different between Table~\ref{table2} and Table~\ref{table1}. It is
because we removed those electrons which have no hadrons in the cone
around them when we calculated the bottom contribution from the
electron spectra for Table~\ref{table2}. This removes different
fractions from $B$ and $D$ decays, which has to be corrected by
simulations. This was done to make the results directly from the
electron spectra comparable to the results from hadron summed
$p_{T}$ histogram fitting. It will be more valuable for the small
acceptance experiments to investigate the $B$ decay contribution
using summed $p_{T}$ histogram fitting.

\begin{center}
\begin{table}
\caption{\label{table2}Fractions of $B$ meson decay contributions
from electron spectra and summed $p_{T}$ histogram fitting in PYTHIA
calculations with parameter set II and modified Peterson
fragmentation function for charm and beauty quarks.}
\begin{tabular}{c|c|c}
\hline
\hline $p_{T}(trig)$ (GeV/c) &From Spectra ($\%$) &From Fitting ($\%$)\\
\hline 2.5 - 3.5 &$8.25 \pm 0.05$ &$8.25 \pm 1.50$ \\
\hline 3.5 - 4.5 &$14.94 \pm 0.13$ &$14.94 \pm 2.54$ \\
\hline 4.5 - 5.5 &$22.60 \pm 0.29$ &$22.60 \pm 4.13$ \\
\hline
\hline
\end{tabular}
\end{table}
\end{center}
\vskip -1.035cm

In conclusion, we found that in order to match the STAR measured
$p_{T}$ shape of $D$ mesons from d+Au collisions at RHIC using the
PYTHIA Monte Carlo generator, a harder charm quark fragmentation
function must be used. Since the $D$ meson production at high
$p_{T}$ is not well constrained by the STAR data and the measurement
of electrons alone has a reduced sensitivity on the $p_{T}$ shape of
$D$ mesons, PYTHIA calculations with parameter set I and set II can
simultaneously describe the $p_{T}$ distributions of $D$ mesons and
non-photonic electrons from semi-leptonic decays of heavy quarks.
These calculations predict that the beauty quark decay electrons are
not a dominant contribution over the entire $p_{T}$ region up to 6-8
GeV/c depending on parameters. The relative contributions to
non-photonic electrons from $B$ and $D$ decays have to be determined
experimentally. We studied the correlations of non-photonic
electrons and inclusive charged hadrons, which can distinguish
between $D$ and $B$ decay contributions to the non-photonic
electrons due to their different decay kinematics. We have proposed
an experimental method to estimate the relative $D$ and $B$ decay
contributions.

\begin{acknowledgements}
The author wishes to thank An Tai, Huan Z. Huang, Lianshou Liu and
Chuck Whitten for their helpful discussions, Ramona Vogt for
suggestions and providing FONLL data. This work was supported in
part by NSFC under project 10575042.
\end{acknowledgements}

{}
\end{document}